\documentstyle[12pt,epsfig]{article}
\begin{document}
\noindent

\begin{center}
{\Large {\bf Gravitational Coupling and the Cosmological Constant}}\\
\vspace{2cm}
 ${\bf Yousef~Bisabr}$\footnote{e-mail:~y-bisabr@srttu.edu.}\\
\vspace{.5cm} {\small{Department of Physics, Shahid Rajaee Teacher
Training University,
Lavizan, Tehran 16788, Iran}}\\
\end{center}
\vspace{1cm}
\begin{abstract}
We deal with a dynamical mechanism in which a large cosmological
constant, as suggested by inflationary scenarios, decays due to
expansion of the universe.  This mechanism has its origin in the
gravitational coupling of the vacuum density.  We assume that the
vacuum couples anomalously to gravity that is the metric tensor that
appears the gravitational part is not the same as that appears the
matter part as suggested by weak equivalence principle. Instead, the
two metric tensors are taken to be conformally related.  We show
that this provides a dynamical mechanism which works during
expansion of the universe. We also consider some observational
consequences of such a gravitational model.

\end{abstract}
\section{Introduction}
There is a large discrepancy between observations and theoretical
estimates on the vacuum energy density.  This problem is known as
the first cosmological constant problem (or the fine tuning problem)
\cite{1} \cite{a1}. The second one known as the coincidence problem
\cite{2} deals with the question that why the vacuum energy is the
same order of magnitude of the matter density.   \\
There have been many attempts trying to resolve the fine tuning
problem \cite{a1}. Most of them are based on the belief that the
cosmological constant may not have such an extremely small value at
all times.  In fact, it has a large value at early times as
suggested by inflationary models and there should exist a dynamical
mechanism working during evolution of the universe which provides a
cancelation of the vacuum energy density at late times \cite{bis}.\\
In this work, we will consider the possibility that this cancelation mechanism is related
to gravitational coupling of matter in a gravitational system. The
matter part of a gravitational system is usually taken to be coupled
with the metric which describes the gravitational part. This is a
normal coupling which has its origin in the weak equivalence
principle. Firstly, it should be pointed out that even
though, observational evidences are fully consistent with weak
equivalence principle, the experiments that support the principle
are recent and it is quite possible that the principle is violated
during evolution of the universe. Secondly, the weak equivalence
principle is a statement about the gravitational coupling of normal
matter and there is no experimental constrains on gravitational
coupling of the other forms of energy densities such as vacuum. \\We will assume that the vacuum
energy density does not couple normally to gravity but rather they
couple abnormally or anomalously, that is the metrics that appear in
the matter and the gravitational parts are different but conformally
related \cite{bis}.  As illustrations, we will consider
such an anomalous gravitational coupling in $f(R)$ gravity \cite{f1} \cite{f2} and Brans-Dicke (BD) theory \cite{bd} and
discuss the conditions that lead to alleviation of the cosmological constant problem.
\section{Field Equations}
 We consider the action functional
 \begin{equation}
S= \int d^{4}x \sqrt{-g}~\{\frac{M_p^2}{2} R -\frac{1}{2}g^{\mu\nu}
\nabla_{\mu} \varphi~ \nabla_{\nu} \varphi -V(\varphi)\}+
S_{m}(A^2(\varphi)g_{\mu\nu} , \psi) \label{b4}\end{equation} where
$M_p^{-2}\equiv 8\pi G$, $G$ is the gravitational constant and
$S_{m}$ is the action of some matter field $\psi$. The function
$A(\varphi)$ is a coupling function that characterizes coupling of
the scalar field $\varphi$ with the matter sector. It defines anomalous gravitational coupling of matter
characterized by the field $\psi$.  In general, such a matter system does not respect the weak equivalence principle. However, the scalar field $\varphi$ can hide the anomalous coupling via chameleon mechanism \cite{cham} and pass local gravity experiments \cite{bc} \cite{ch}.\\
Variation of this
action with respect to the metric tensor $g^{\mu\nu}$, gives
\begin{equation}
G_{\mu\nu}=M_p^{-2}(T^{\varphi}_{\mu\nu}+
T^{m}_{\mu\nu}) \label{b6}
\label{b7}\end{equation} where
\begin{equation}
T^{\varphi}_{\mu\nu}=\nabla_{\mu}\varphi
\nabla_{\nu}\varphi-\frac{1}{2}g_{\mu\nu}\nabla^{\gamma}\varphi
\nabla_{\gamma}\varphi-V(\varphi)g_{\mu\nu}
\label{b8}\end{equation}
\begin{equation}
T^m_{\mu\nu}=\frac{-2}{\sqrt{-g}}\frac{\delta S_{m}}{\delta
g^{\mu\nu}} \label{b9}\end{equation} are stress-tensors of the
scalar field and the matter system. If we vary the action with respect to the
scalar field $\varphi$, we obtain
\begin{equation}
\Box\varphi-\frac{dV(\varphi)}{d\varphi}=-\frac{\beta(\varphi)}{M_p}
T^{m} \label{b11}\end{equation} or, equivalently,
\begin{equation}
\nabla^{\mu}T^{\varphi}_{\mu\nu}=-\frac{\beta(\varphi)}{M_p}
\nabla_{\nu}\varphi T^m \label{b8-10}\end{equation}
 where
\begin{equation}
\beta(\varphi)=M_p \frac{d\ln A(\varphi)}{d\varphi}
\label{b8-1}\end{equation} and $T^m\equiv g^{\mu\nu}T^m_{\mu\nu}$.
The two stress tensors $T^m_{\mu\nu}$ and
$T^{\varphi}_{\mu\nu}$ are not separately conserved due to coupling
of the scalar field $\varphi$ with matter.  This can be seen by applying the
Bianchi identities $\nabla^{\mu} G_{\mu\nu}=0$ to (\ref{b6}) which results in
\begin{equation}
\nabla^{\mu}T^{m}_{\mu\nu}=-\nabla^{\mu}T^{\varphi}_{\mu\nu}=
\frac{\beta(\varphi)}{M_p}\nabla_{\nu}\varphi~T^{m}\label{b13}\end{equation}
The parameter $\beta$ is generally a function of time. We will, however, restrict ourselves to
the case that it can be regarded as a constant parameter. There are two important cases
in which the function $\beta(\varphi)$ takes a constant configuration; $f(R)$ gravity and Brans-Dicke
theory.
~~~~~~~~~~~~~~~~~~~~~~~~~~~~~~~~~~~~~~~~~~~~~~~~~~~~~~~~~~~~~~~~~~~~~~~~~~~~~~~~~~~~~~~~~~~~~~~~~~~~~~~~~~~~~~~~~~~~~~~~~~~~~~~~
\subsection{$f(R)$ Gravity}
The action for an $f(R)$ gravity theory in Jordan frame is given by \cite{f2}
\begin{equation}
S_{JF}= \frac{1}{2}M_p^2\int d^{4}x \sqrt{-\bar{g}} f(\bar{R})
+S_{m}(\bar{g}_{\mu\nu}, \psi)\label{b1}\end{equation} where
$\bar{g}_{\mu\nu}$ is the metric in Jordan frame.  A conformal
transformation
\begin{equation}
g_{\mu\nu} =A^{-2}(\varphi)~ \bar{g}_{\mu\nu}
\label{b2}\end{equation} with
$A^{-2}(\varphi)\equiv\frac{df}{dR}=f^{'}(R)$  together with
\begin{equation} \varphi = \frac{M_p}{\beta } \ln A(\varphi)
\label{b3}\end{equation} and $\beta=-\sqrt{\frac{1}{6}}$, transforms
(\ref{b1}) into the action (\ref{b4}) with a potential \cite{soko}
\cite{w}
\begin{equation}
V(\varphi(R))=\frac{M_p^2}{2}(\frac{R}{f'(R)}-\frac{f(R)}{f'^2(R)})
\label{b5}\end{equation}
~~~~~~~~~~~~~~~~~~~~~~~~~~~~~~~~~~~~~~~~~~~~~~~~~~~~~~~~~~~~~~~~~~~~~~~~~~~~~~~~~~~~~~~~~~~~~~~~~~~~~~~~~~~~~~
\subsection{Scalar-Tensor Gravity}
The general action of a scalar-tensor gravity is given by \cite{pol}
\begin{equation}
S_{JF}=\frac{1}{16\pi G} \int d^{4}x
\sqrt{-\bar{g}}~\{F(\phi)\bar{R}-Z(\phi)\bar{g}^{\mu\nu}
\bar{\nabla}_{\mu} \phi~ \bar{\nabla}_{\nu} \phi -2U(\phi)\}+ S_{m}(
\bar{g}_{\mu\nu}, \psi_m) \label{d1}\end{equation} where $F(\phi)$,
$Z(\phi)$ and $U(\phi)$ are some functions\footnote{ One can always
redefine the scalar field to reduce $F(\phi)$ and $Z(\phi)$ to only
one unknown function.}.  This action is reduced to the action
(\ref{b4}) by the conformal transformation (\ref{b2}) with
$A(\varphi)=F^{-1/2}(\phi)$ and
\begin{equation}
(\frac{d\varphi}{d\phi})^2=2M_p^2[\frac{3}{4}(\frac{d\ln
F(\phi)}{d\phi})^2+\frac{Z(\phi)}{2F(\phi)}]
\label{d2}\end{equation}
\begin{equation}
V(\varphi)=M_p^2 U(\phi)F^{-2}(\phi)\label{d3}\end{equation} It is cleat from (\ref{d2}) that the
coupling function $A(\varphi)$ depends on the functions $F(\phi)$,
$Z(\phi)$ and $U(\phi)$ through the relation (\ref{d2}).  For some
particular choices of these functions, $\beta$ takes a constant
configuration and then, as a result of (\ref{b8-1}), $A(\varphi)$
takes an exponential form. This defines a class of scalar-tensor
theories and we restrict ourselves to this class. One
important theory in this class is given by BD parameterization in
which $F(\phi)=16\pi G\phi$, $Z(\phi)=16\pi G\omega_{BD}/\phi$ and
$U(\phi)=8\pi G W(\phi)$, and then
\begin{equation}
S_{JF}= \int d^{4}x \sqrt{-\bar{g}} (\phi \bar{R}
-\frac{\omega_{BD}}{\phi}\bar{g}^{\mu\nu}\bar{\nabla}_{\mu}\phi
\bar{\nabla}_{\nu}\phi-W(\phi))+S_{m}(\bar{g}_{\mu\nu},
\psi)\label{d5}\end{equation} with $\omega_{BD}$ and $W(\phi)$ being
BD parameter and the potential in Jordan frame, respectively.  This
action is reduced to (\ref{b4}) by \cite{bis} \cite{sc}
\begin{equation}
A(\varphi)=e^{\beta_{BD}\varphi/M_p}\label{d6}\end{equation}
\begin{equation}
\varphi(\phi)/M_p=\sqrt{\omega_{BD}+3/2}\ln (\frac{\phi}{\phi_0})
\label{a3}\end{equation}
\begin{equation}
V(\varphi)=W(\phi(\varphi))~e^{8\beta_{BD}\varphi/M_p}
\label{d6-1}\end{equation} where $\phi_0\sim G^{-1}$ and
$\beta_{BD}=-1/2\sqrt{\omega_{BD}+3/2}$.  When $\omega_{BD}\rightarrow 0$, then
$\beta_{BD}\rightarrow \frac{-1}{\sqrt{6}}$ and Einstein frame
representations of BD model and $f(R)$ gravity are the same.
~~~~~~~~~~~~~~~~~~~~~~~~~~~~~~~~~~~~~~~~~~~~~~~~~~~~~~~
\section{Cosmological Setting}
We use a spatially flat
 homogeneous and isotropic cosmology described by Friedman-Robertson-Walker
spacetime
\begin{equation}
ds^2=-dt^2+a^2(t)(dx^2+dy^2+dz^2)
\end{equation}
where $a(t)$ is the scale factor. We also take
$T^m_{\mu\nu}$ as the stress-tensor of a perfect fluid with energy density
$\rho_{m}$ and pressure
$p_m$.  The Friedman equation is
\begin{equation}
3H^2=M_p^{-2}(\rho_{m}+\rho_{\varphi})
\label{b14}\end{equation}
 where
$H=\frac{\dot{a}}{a}$, $\rho_{\varphi}=\frac{1}{2}\dot{\varphi}^2+V(\varphi)$
and overdot
indicates differentiation with respect to time $t$. From
(\ref{b11}) and (\ref{b13}), we obtain
\begin{equation}
\ddot{\varphi}+3\frac{\dot{a}}{a}\dot{\varphi}+\frac{dV(\varphi)}{d\varphi}=-\frac{\beta(\varphi)}{M_p}
(\rho_{m}-3p_m) \label{b15}\end{equation}
\begin{equation}
\dot{\rho}_{m}+3\frac{\dot{a}}{a}(\omega_m+1)\rho_{m}=Q \label{b17}\end{equation}
\begin{equation}
\dot{\rho}_{\varphi}+3\frac{\dot{a}}{a}(\omega_{\varphi}+1)\rho_{\varphi}=-Q
\label{b18}\end{equation} where
\begin{equation}
Q=\frac{\beta(\varphi)}{M_p} \dot{\varphi}(\rho_{m}-3p_m)
\label{b-18}\end{equation} is the interaction term, $\omega_m\equiv
p_m/\rho_m$ and $\omega_{\varphi}\equiv p_{\varphi}/\rho_{\varphi}$.
The direction of energy transfer depends on the sign of $Q$.  For
$Q>0$, the energy transfer is from the scalar field to the matter
system and for $Q<0$, the reverse is true. The solution of equation (\ref{b17})
is
\begin{equation}
\rho_m=\rho_{0m}~a^{-3(\omega_m+1)}~e^{\frac{(1-3\omega_m)}{M_p}\int
\beta d\varphi} \label{c1}\end{equation} in which $\rho_{0m}$ is an
integration constant. This solution indicates that the evolution of
energy density is modified due to interaction of $\varphi$ with
matter. The expression (\ref{c1}) can be written as
\begin{equation}
\rho_m=\rho_{0m}~a^{-3(\omega_m+1)+\epsilon}
\label{c3}\end{equation}
with $\epsilon$ being defined by
\begin{equation}
\epsilon=\frac{(1-3\omega_m)\int\beta d\varphi}{M_p\ln a}
\label{c2-1}\end{equation} For constant $\beta$, (\ref{c2-1}) reduces to
\begin{equation}
\varphi=\sigma M_p\ln a \label{c2}\end{equation} with $\sigma$ being
a constant defined by the relation $\epsilon=\beta
\sigma(1-3\omega_m)$.  Here the parameter $\beta$ takes $\beta_{R}$ and $\beta_{BD}$
in $f(R)$ gravity and BD theory, respectively.\\
 The expression (\ref{c3}) states that when
$\epsilon>0$, matter is created and energy is constantly injecting
into the matter so that the latter will dilute more slowly compared
to its standard evolution $\rho_m\propto a^{-3(\omega_m+1)}$.
Similarly, when $\epsilon<0$ the reverse is true, namely that matter
is annihilated and the direction of energy transfer is outside of
the matter system so that the rate of dilution is faster than
the standard one.\\
Let us first take the matter system to be vacuum energy density characterized by a perfect
fluid
with equation of state parameter $\omega_m=-1$.  Thus the vacuum energy density is
anomalously coupled to gravity.  In this case, (\ref{c3}) becomes $\rho_{\Lambda}\equiv\rho_m=\rho_{0\Lambda}~a^{\epsilon}$
 where we have set $\rho_{0m}=\rho_{0\Lambda}$.  In an expanding universe, the
requirement that a large vacuum energy density $\rho_{\Lambda 0}$
reduces during the expansion needs $\varepsilon<0$.  Note that in this case the direction of energy transfer is
out of vacuum. We can use Friedman equation (\ref{b14}) and the equations (\ref{b17}) and (\ref{b18}) to write
the deceleration parameter
\begin{equation}
q(z)=-1-\frac{\dot{H}}{H^2}=-1-\frac{\frac{1}{2}[\epsilon\Omega_{\Lambda}-\frac{Q}{H\rho_c}-3(\omega_{\varphi}+1)\omega_{\varphi}]}
{(\Omega_{\Lambda}+\Omega_{\varphi})}\label{cd7}\end{equation}
where $\Omega_{\Lambda}=\rho_{\Lambda}/\rho_c$, $\Omega_{\varphi}=\rho_{\varphi}/\rho_c$ and $\rho_c=3H^2_0 M_p^2$ is the critical
density.  On can use (\ref{b-18}) to show that $\frac{Q}{H\rho_c}=\epsilon\Omega_{\Lambda}$.  This together with spatial flatness,
which requires that $\Omega_{\Lambda}+\Omega_{\varphi}=1$, reduce (\ref{cd7}) to
\begin{equation}
q(z)=-1+\frac{3}{2}(\omega_{\varphi}+1)(1-\Omega_{\Lambda})\label{dd7}\end{equation}
which accelerating expansion requires that $\omega_{\varphi}+1<\frac{2}{3(1-\Omega_{\Lambda})}$.\\
The coincidence problem concerns with the constancy of the ratio $r=\frac{\rho_m}{\rho_c}$. Using (\ref{b17}) and (\ref{b18}), we
can write
\begin{equation}
\dot{r}=rH[\epsilon(r+1)+(\omega_{\varphi}-\omega_m)]\label{dd7}\end{equation}
where we have used $Q=\epsilon H \rho_m$.  If we set $\omega_m=0$, the constancy of $r$ requires that
$\omega_{\varphi}= -\frac{1}{3}\epsilon (r+1)$ which $\omega_{\varphi}>1$ for a vacuum decay ($\epsilon<0$).

~~~~~~~~~~~~~~~~~~~~~~~~~~~~~~~~~~~~~~~~~~~~~~~~~~~~~~~~~~~~~~~~~~~~~~~~~~~~~~~~~~~~~~~~~~~~~~~~~~~~~~~~~~~~

\section{Conclusion}
There is no observational constraints on gravitational coupling of
matter systems during expansion of the universe such as
baryons, radiation, vacuum, dark matter or dark energy. The weak
equivalence principle which is supported by recent observations
\cite{will} only constrains the baryons. Thus it is quite possible
that other kinds of matter
systems such as vacuum couples differently with gravity.\\
We consider a dynamical mechanism which works with expansion of the
universe and has its origin in gravitational coupling of the vacuum.
We assume that vacuum couples anomalously with gravity in the sense
that the metrics of the gravitational and matter parts are not the
same but conformally related. The conformal factor is then
controlled by a dynamical scalar field.  We provide some examples
from $f(R)$ gravity and
BD theories.  We have shown that the cosmological constant problem may be alleviated in such theories due to anomalous gravitational
coupling of vacuum.

~~~~~~~~~~~~~~~~~~~~~~~~~~~~~~~~~~~~~~~~~~~~~~~~~~~~~~~~~~~~~~~~~~~~~~~~~~~~~~~~~~~~~~~

\end{document}